# Electrically Reconfigurable Intelligent Optoelectronics in 2-D van der Waals Materials

Yu Wang[a,b], Dehui Zhang[c], Yihao Song[d], Jea Jung Lee[d], Meng Tian[a,b], Souvik Biswas[e,*], Fengnian Xia[d,*] and Qiushi Guo[a,b,*]

[a]*Photonics Initiative, Advanced Science Research Center, City University of New York, 85 St Nicholas Terrace, New York, 10031, NY, USA*
[b]*Physics Program, Graduate Center, City University of New York, 365 5th Ave, New York, 10016, NY, USA*
[c]*Department of Electrical Engineering and Computer Sciences, University of California, Berkeley, 566 Cory Hall, Berkeley, 94720, CA, USA*
[d]*Department of Electrical Engineering, Yale University, 15 Prospect St, New Haven, 06511, CT, USA*
[e]*E. L. Ginzton Laboratory, Stanford University, 348 Via Pueblo Mall, Stanford, 94305, CA, USA*

ABSTRACT

In optoelectronics, achieving electrical reconfigurability is crucial as it enables the encoding, decoding, manipulating, and processing of information carried by light. In recent years, two-dimensional van der Waals (2-D vdW) materials have emerged as promising platforms for realizing reconfigurable optoelectronic devices. Compared to materials with bulk crystalline lattice, 2-D vdW materials offer superior electrical reconfigurability due to high surface-to-volume ratio, quantum confinement, reduced dielectric screening effect, and strong dipole resonances. Additionally, their unique band structures and associated topology and quantum geometry provide novel tuning capabilities. This review article seeks to establish a connection between the fundamental physics underlying reconfigurable optoelectronics in 2-D materials and their burgeoning applications in intelligent optoelectronics. We first survey various electrically reconfigurable properties of 2-D vdW materials and the underlying tuning mechanisms. Then we highlight the emerging applications of such devices, including dynamic intensity, phase and polarization control, and intelligent sensing. Finally, we discuss the opportunities for future advancements in this field.

## 1. Introduction

As evidenced by the success of fiber-optical telecommunications for decades, light is an excellent information carrier, capable of conveying information through its amplitude, phase, polarization, and wavelengths. Compared to electronics, photonic information processing and sensing enjoy numerous advantages such as high transmission speed, long-distance propagation, high data rates[1], resilience to electromagnetic interference, ease of wavelength-[2, 3], time-[4], space-[5, 6] and polarization-division multiplexing[7], and high coherence[8]. As we are transitioning to a new era of the information age driven by mobile devices, intelligent algorithms, and increasingly more computing resources along with burgeoning quantities of data, we also confront many outstanding technological challenges and scientific inquiries in optoelectronics. For instance, can we encode more information into light? How can we extract higher-dimensional information from light? How to process the information carried by light in a more energy-efficient manner? And, how can we make photonic sensors and information processors more compact?

Intertwined with these inquiries lies a pivotal characteristic of optoelectronics – reconfigurability. Over time, we have been extensively exploiting the reconfigurability of optoelectronic devices to encode, manipulate, and sense the diverse information carried by light. For instance, reconfigurable electro-optic crystals such as lithium niobate have been widely employed to encode data into the phase or intensity of light for communication applications[9]; electrically or mechanically tunable mirrors and gratings play a crucial role in light detection and ranging (LIDAR)[10] as well as spectroscopy[11]. However, many existing reconfigurable optoelectronic devices are still bulky, mono-functional, and exhibit slow response speeds, which limit their applicability in various fields. Furthermore, compared to mechanically or magnetically tunable devices, which typically involve moving parts and magnets, electrically reconfigurable optoelectronic devices present clear advantages in terms of reduced noise and enhanced integration

*Corresponding authors

souvik@stanford.edu (S. Biswas); fengnian.xia@yale.edu (F. Xia); qguo@gc.cuny.edu (Q. Guo)
ORCID(s):

capabilities. Looking forward, advancements in optoelectronic technologies will benefit greatly from the discovery of new materials with strong and fast electrical optoelectronic reconfigurability, alongside innovative device physics and device architectures that enable multi-functionality.

In the quest for new reconfigurable optoelectronic materials, 2-D vdW materials and their heterostructures stand out as promising candidates. Their inherent thinness not only grants them high responsiveness to external stimuli, but also endows them with a range of intriguing optoelectronic properties uncommon in conventional bulk crystal lattice materials. These include the inherently high surface-to-volume ratio, reduced carrier screening[12, 13], quantum confinement[12, 14, 15], and the ability to construct tailored band-structures with interesting quantum geometric and topological properties via assembly and twisting of vdW layers[16–22]. More interestingly, the interplay of these physical properties can give rise to exotic, tunable optoelectronic properties. For instance, the reduced carrier screening effects in 2-D semiconductors or semi-metals enable strong room-temperature dipole resonances, which can be highly tunable via carrier density tuning[23, 24]. The quantum confinement exhibited by 2-D semiconductors enables various mechanisms for efficiently tuning their optical transition edges and bandgaps[14, 25–28]. Additionally, when coupled with bandgap tuning, the intricate band structure of twisted vdW materials can be exploited to manipulate the material's topological properties and response to light[29]. Finally, from a device and system viewpoint, thanks to the naturally passivated surface, 2-D vdW materials are amenable to integrating with optical microstructures such as cavities[30–33], waveguides[34–38], and gratings without incurring the lattice mismatch or deep-level trapping centers, thus circumventing a common challenge observed in heterostructures made of two different bulk crystalline materials.

In recent decades, the remarkable electrical reconfigurability of 2-D materials has been extensively explored across various optoelectronic device applications, such as integrated optical modulators[34, 35, 39, 40], tunable chip-scale classical or quantum light sources[28, 41–44]. Meanwhile, recent breakthroughs in material discovery and intelligent learning algorithms have catalyzed a transformation in optoelectronics research, turning it into a highly interdisciplinary domain encompassing devices and materials engineering, condensed matter physics, and machine learning[45]. This review article seeks to establish a connection between the fundamental physics underlying reconfigurable optoelectronics in 2-D materials and their burgeoning applications in intelligent optoelectronics.

## 2. Electrically tunable optoelectronic properties in 2-D materials

In Fig.1, we highlight several optoelectronic properties that can be electrically tuned in 2-D vdW materials, such as the complex refractive index, birefringence, photo-responsivity, spectral and polarization response. We discuss the underlying physical principles that facilitate the efficient tuning of these properties. Furthermore, we establish connections between these tuning mechanisms and their potential applications.

**Complex refractive index tuning:** In addition to the Kerr effect and Pockels effect, various electro-optic effects—where an applied external electric field alters the real or imaginary components of a material's refractive index—exist in low-dimensional materials and have been leveraged in device applications such as optical modulators. For instance, the Quantum Confined Franz–Keldysh (QCFK) effect occurs when an electric field tilts the semiconductor band edges, altering the shape of the wavefunction. This change allows the tail of the wavefunction to extend into the bandgap, thereby increasing the absorption of photons with energies below the bandgap[25, 26, 46–48]. Another mechanism is known as the field-induced Pauli-blocked Burstein–Moss (BM) shift[25, 48], which is usually observed in chemically doped narrow-band gap semiconductor materials. As the charge carrier density increases and the Fermi level moves into the conduction or valence band, fewer unoccupied states remain available for optical transitions, resulting in reduced absorption.

Realizing dynamic light-field control in optoelectronic devices requires physical mechanisms that can largely modulate the light field (amplitude or phase) in a short interaction length or device footprint. This necessitates a mechanism capable of significantly altering the refractive index of a material, a challenging task for bulk materials[49]. As illustrated in Fig. Fig. 1a, due to their atomically thin nature and reduced screening, 2-D materials host strong dipole-like excitations at room temperature, such as excitons and plasmons. These dipole-like excitations can greatly enhance the strength of light-matter interaction, leading to enhanced reflection or absorption. Moreover, these 2-D dipole resonances are highly tunable via electrostatic gating. Typically, near these dipole resonances, the material is dispersive and the real and imaginary parts of the material's refractive index undergo dramatic changes. In the visible and near-IR frequency range, the complex refractive index of 2-D semiconductors near exciton resonances can be well approximated by a Lorentzian function which is described by oscillator strength, resonance frequency, and linewidth (which is dominated by radiative, non-radiative and dephasing rates of excitons). Through charge injection via electrical gating, the resonance frequency, oscillator strength, and linewidth can all be significantly altered. This, in turn, strongly modulates the Lorentzian function and the amplitude and phase of reflected and transmitted light. Compared to other electro-optic properties in 2-D semiconductors such as field-induced BM shift and QCFK effect [25], the complex refractive index modulation around the exciton resonances is much stronger, which can even exceed unity[50]. A similar principle can also be applied to mid-IR and THz plasmon resonances in 2-D semi-metals such as graphene[51–54] or highly doped 2-D semiconductors such as black phosphorus (BP)[55].

| Tunable properties | Physical mechanisms | Illustration | Applications |
|---|---|---|---|
| (a) Complex refractive index tuning | Carrier density + 2-D dipolar resonances (excitons, plasmons) | | Amplitude/phase modulators (switches), beam steering, free-space communication, dielectric sensing |
| (b) Birefringence tuning | Carrier density + anisotropic 2-D dipolar resonances | | Tunable waveplate, polarization encoding and sensing, Stokes polarimetric detection |
| (c) Photo-responsivity tuning | Photocarrier lifetime tuning, Band alignment tuning | | In-sensor matrix multiplication, in-sensor computing, in-memory computing |
| (d) Band-structure tuning | Quantum confined Stark effect, Franz-Keldysh effect, inversion symmetry breaking | | Wavelength tunable light emitters, single-pixel spectrometers, hyperspectral imaging |
| (e) Quantum geometry tuning | Tuning of the elements of nonlinear conductivity tensor | | Single-pixel stokes polarimetric detection and spectrometry |

**Figure 1:** Electrically-tunable optoelectronic properties in 2-D materials and their representative applications. (a) The complex refractive index of 2-D materials can be substantially adjusted around dipole resonances (such as excitons and plasmons) through carrier density modulation. Modulating the real part of the refractive index will introduce phase modulation, whereas tuning the imaginary part will lead to amplitude modulation of light interacting with 2-D materials. (b) Tuning the birefringence properties of 2-D materials can be accomplished by tuning the anisotropic dipole resonances with carrier density. The tunable birefringence has significant implications in encoding information on light polarization and its detection. (c) The photo-responsivity of photodetectors based on 2-D materials can be modulated via carrier density modulation, offering implications for image sensors with neuromorphic computing capabilities. (d) The band structure of 2-D materials can be tuned by a vertical displacement field due to the quantum-confined Stark effect (QCSE) and inversion symmetry breaking. The capability of bandgap tuning in 2-D materials can enable wavelength-tunable light sources, ultra-compact single-pixel spectrometers, and hyperspectral imagers. (e) The quantum geometric characteristics of Bloch bands, including Berry curvature and quantum metric can be tuned by a vertical displacement field[29], which plays important roles in simultaneously analyzing the spectral and polarization information of light.

**Birefringence tuning:** As detailed above, 2-D materials can enable large refractive index tuning through dipole resonance modulation via charge injection or electrostatic gating. For highly anisotropic 2-D materials, this can have interesting implications in the polarization domain, as illustrated in Fig. 1b. For example, BP, a prototypical strongly birefringent material, hosts stable excitons that are quasi-1D in nature - polarized along the armchair direction (optically active axis)[56–58]. The zigzag direction (optically inactive axis) behaves like a minimally dispersive semiconductor. As a result, in the presence of charge injection or electrostatic gating, a similar refractive index tuning (as mentioned above) takes place along the optically active crystal orientation, whereas the inactive axis remains passive. This results in strong anisotropic amplitude and phase modulation of the incident light[59]. Such an anisotropic modulation can lead to the generation of a variety of polarization states on the Poincaré sphere as well as strong modulation of incoming polarization. Furthermore, combined with the photo-response of the materials, it can lead to complete Stokes polarimetric detection. Material choices for polarization tuning are discussed in Section 3.3.

**Photo-responsivity tuning:** At the heart of any optoelectronics application lie photodetectors – devices that convert light into electrical signals. The key parameter governing a photodetector's performance is its photo-responsivity ($R$), which represents the ratio between the generated photocurrent ($I_{ph}$) or photovoltage ($V_{ph}$) and the incident light power ($P$), commonly expressed in A/W or V/W. $R$ is usually influenced by various intrinsic factors of the semiconductor photodetector material itself, such as bandgap, carrier lifetime, and band alignment. Additionally, extrinsic factors such as the wavelength and polarization of incident light, as well as the incident light power, may also affect $R$. As shown in Fig. 1c, in 2-D materials and their heterostructures, tuning of $R$ can be readily achieved by adjusting the doping concentration via electrostatic gating of the channel of field-effect transistor-like photodetectors[60–63] or by locally electrostatic gating of homogeneous p-n junctions[64–66]. In addition, tuning of $R$ can also be accomplished by adjusting the band alignment through the global gating of heterogeneous p-n junctions[67, 68]. In these device configurations, gate voltage governs responsivity through the combination of two mechanisms. First, it regulates the carrier concentration in 2-D material without light excitation, and the lifetime of photogenerated carriers can be tuned because the recombination rate depends on the carrier concentration. Second, it controls the carrier collection efficiency by tuning the band alignment along the device channel. It is important to note that the programming of $R$ can also be non-volatile by leveraging gate dielectrics with charge-storing mechanisms[61, 62, 65, 69], or by introducing conductive ions and controlling their migration with voltage pulses[70]. The ability to dynamically program and store the value of $R$ in 2-D materials enables the development of emerging applications such as image sensors with in-sensor computing capabilities[61, 62, 64, 71].

**Band structure tuning:** As a fundamental property of a material, the band structure governs the optical transition energy and the strength of the optical transition dipole and light-matter interactions. In traditional bulk semiconductors, the bandgap is determined by the chemical composition and specific arrangement of crystal lattices, making it challenging to tune. However, in 2-D materials, where each atomic layer is bonded by weak van der Waals forces rather than chemical bonds, each layer behaves like a quantum well, exhibiting strong quantum confinement. In such a system, the electric field can be leveraged to effectively alter the material's bandgap, resembling the quantum-confined Stark effect (QCSE) observed in III-V semiconductor quantum wells[72]. Moreover, due to reduced electric-field screening in 2-D materials, such a bandgap modulation based on QCSE can be more pronounced. As shown in Fig.1d, a common method of efficiently modulating the bandgap in 2-D materials based on QCSE involves applying a vertical displacement field across the material by applying both top and bottom gate voltages (dual-gate configuration). Experimental realization of bandgap tuning in various 2-D materials using dual-gate configuration has been demonstrated. For example, for bilayer $MoS_2$, the optical bandgap (exciton transition energy) can be tuned from 1.85 eV to 1.55 eV using a displacement field of 1.2 V/nm, corresponding to a tuning rate of ~275 meV per 1 V/nm[73]. In BP with a thickness of around 10 nm, the bandgap can be continuously tuned from approximately 300 to below 50 meV using a moderate displacement field of around 1 V/nm[27], significantly expanding BP's spectral response from the mid-wave infrared (MWIR) regime to the long-wave infrared (LWIR) regime[28].

As a different bandgap tuning mechanism, the vertical displacement field can also open a bandgap by breaking the inversion symmetry. This mechanism is crucial for tuning the bandgap of 2-D semi-metals whose intrinsic bandgap is zero. Experimental[74–77] and theoretical[78, 79] studies have demonstrated that the band structure of Bernal-stacking bilayer graphene can be continuously modified by electric fields, converting the zero bandgap material into a semiconductor with a bandgap of up to approximately 300 meV. Efficiently and continuously varying the band structure using electrical methods offers new avenues for modulating multiple physical properties of materials, including spectral response, light emission wavelength, quantum efficiency, photo-responsivity, refractive index, and more, which can find applications in developing compact spectrometers[80, 81], hyperspectral imagers[82], and wavelength-tunable light sources[83–85].

**Quantum geometry tuning:** The quantum geometric characteristics of Bloch wave functions, including Berry curvature and quantum metric, are fundamental in shaping the behavior of electrons in their ground state, leading to various condensed matter phenomena such as the electric polarization of crystals[86], orbital magnetization[87], and quantum and anomalous Hall effects[88]. Moreover, it has been discovered that the nonlinear light-matter interactions are also largely governed by the quantum geometric properties of the materials[89–91]. A notable example is the bulk photovoltaic effect (BPVE) observed in inversion-symmetry broken materials, wherein excited charge carriers can spontaneously flow through the bulk of the material even in the absence of applied electric fields[92]. Such bulk geometric photocurrents due to BPVE are particularly sensitive to crystal symmetry and light polarization configurations. In 2-D materials and their heterostructures such as twisted double bilayer graphene (TDBG), it has been shown that Berry curvature is highly tunable via the applied out-of-plane displacement field[29], as shown in Fig. 1e. Such a reconfigurable quantum geometry can be employed to analyze incident light polarization configurations.

## 3. Dynamic light field modulation and beam steering

The conventional approach to controlling light relies on the use of bulky and expensive optical components. In recent years, we have witnessed remarkable advancements in manipulating the flow of light using ultra-thin, lightweight flat optical devices, such as metasurfaces[93, 94]. The next frontier in flat optics lies in achieving programmable, fast, and dynamic control over various properties of light fields (intensity, phase, and polarization), a capability crucial for various emerging applications such as free-space optical communications, optical neural networks, LIDAR, computational imaging and sensing[95], and dynamic holography for augmented and virtual reality-based technologies[96]. However, realizing dynamic control of light fields in flat optical devices, which are usually composed of a dense array of dielectric resonators or metal plasmonic resonators, remains an outstanding challenge. This is largely due to the inherently weak electro-absorption and electro-refraction effects in conventional metals and semiconductors. Therefore, the pursuit of new materials and new mechanisms that can be harnessed to construct versatile and tunable flat optical devices holds immense value.

One promising approach to achieving a strong modulation of the phase and intensity of the light field involves the modulation of the material's optical properties around its exciton resonances. In earlier studies on exciton resonances in III-V quantum wells, it has been shown that the resulting refractive index changes induced by the QCSE can be quite substantial, in the range of $\Delta n \sim 0.01$ to $0.04$ [14, 97, 98]. However, in conventional bulk semiconductors, the exciton binding energy is exceedingly small (typically in the range of 5–10 meV) due to strong self-screening effects[99], precluding the existence of appreciable excitonic features at room temperature. In contrast, 2-D semiconductors such as monolayer transition metal dichalcogenides (TMDCs) host room-temperature exciton resonances due to their reduced dimensionality and hence reduced screening effects and large exciton binding energies (often in the hundreds of meV)[100]. It has been shown that near unity reflectivity modulation can be achieved by modulating exciton resonances in monolayer TMDCs[101, 102]. Furthermore, unlike transparent conducting oxides, which typically enable tunable amplitude and phase modulation of light in the near-infrared spectrum[103–105], exciton resonances in 2-D materials primarily occur at visible frequencies. This characteristic significantly broadens the possibilities within the visible spectrum, unlocking new avenues for applications in digital holography, imaging, cloaking, and virtual reality as well as in atomic physics for manipulation of optical tweezers.

### 3.1. Active light field control

Van de Groep et al. demonstrated the first atomically thin $WS_2$ zone plate lens, showing that excitonic resonances in monolayer 2-D materials can be engineered to dynamically manipulate wavefront and the focusing efficiency[106]. As shown in Fig. 2a, in their device, large scale, Chemical vapor deposition (CVD)-grown monolayer $WS_2$ is patterned into 202 concentric rings, similar to conventional zone plates. When incident light aligns with $WS_2$'s exciton resonance wavelength around 625 nm, it experiences diffraction by the $WS_2$ concentric rings, resulting in constructive interference and the creation of a sharp focus. Remarkably, the focusing efficiency of the $WS_2$ zone plate (defined as the ratio of the integrated intensity in focus to the integrated incident intensity on the zone plate) can be dynamically controlled by modulating $WS_2$'s exciton resonances through the ionic liquid gating. Specifically, as a result of applying 3 V gate bias, the focusing efficiency is modulated by ~33% at $\lambda = 625$ nm, despite the single-pass interaction of light with the atomically thin structure (Fig. 2b). In addition, the dynamic tuning currently exhibits a response time of $39 \pm 3$ ms (rise time) and 16 ± 1 ms (fall time) due to ion-transport-limited complex formation and disassembly of the ionic-liquid electrical double layer. It has been shown by Datta et al. that a similar gating scheme can achieve ~330 MHz modulation speed[39], inciting confidence that future work on solid-state electrostatic gating can be explored to further improve the modulation speed. This work has introduced a prototype of dynamically controllable 2-D optical devices, serving as a paradigm for the development of more advanced capabilities in wavefront manipulation.

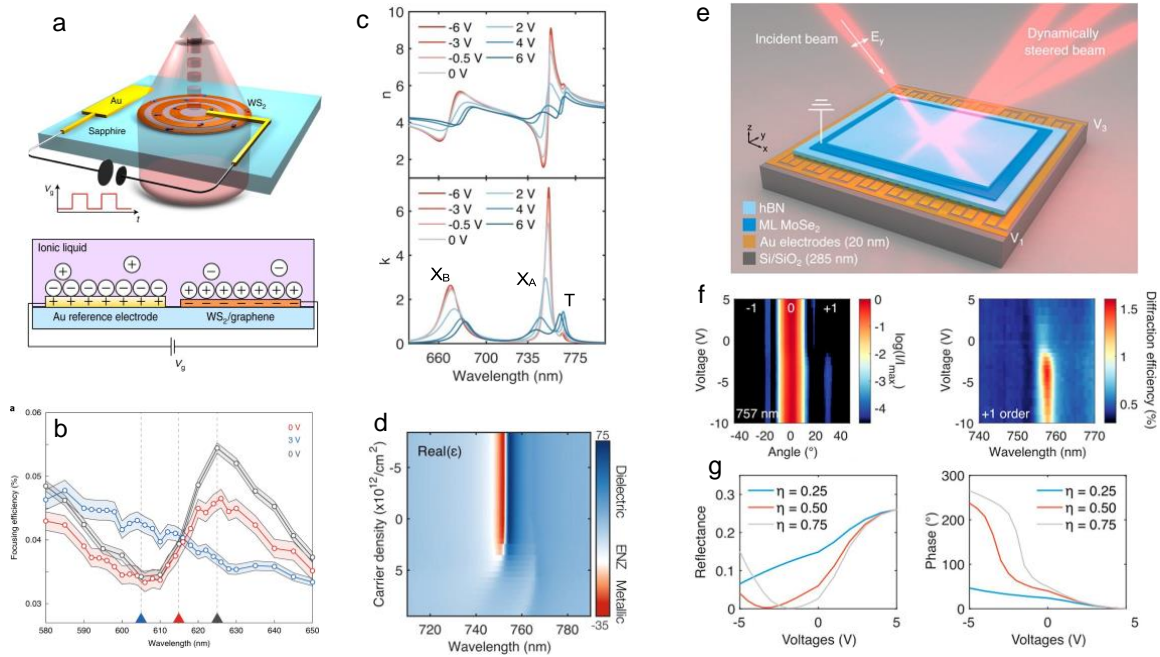

**Figure 2:** (a) Schematic representation of single-layer Fresnel zone-plate WS$_2$ lens in an electrochemical cell. The dynamic ionic-liquid gate bias leads to the modulation of focusing efficiency by quenching exciton resonances. The gating scheme is shown in the schematic below. (b) Focusing efficiency spectra of single layer WS$_2$ lens for three different bias conditions: the initial state without gating ($V_g$ = 0 V (red)), the biased state ($V_g$ = 3 V (blue)), and the final state ($V_g$ = 0 V (grey)) which displays hysteretic gating behavior. One standard deviation is shown in the shaded regions. (c) Spectra of the real and imaginary parts of the complex refractive index of MoSe$_2$ for different gate voltages (with the excitonic resonances labeled) at a sample temperature of 4 K. (d) Spectra of the real part of complex dielectric permittivity as a function of carrier density. In the vicinity of A exciton resonance (754 nm), the MoSe$_2$ undergoes a transition from being optically metallic to optically dielectric. (e) Illustration and device structure of the tunable excitonic metasurface for dynamic beam steering. (f) Left: Measured diffraction pattern at the A excitonic resonance (754 nm) as a function of bias voltage. Right: Electrically tuning of the diffraction efficiency of the +1 order beam as a function of bias voltage. (g) Voltage-dependence of reflectance and phase at the A exciton wavelength of 757 nm for different radiative efficiencies showing an improvement in the performance for $\eta \geq 0.5$. (a–b) are reprinted with permission from Ref.[106], copyright (2020) Springer Nature Limited. (c–d) are reprinted with permission from Ref.[50], copyright (2021) American Chemical Society. (e-g) are reprinted with permission from Ref.[107], copyright (2023) American Chemical Society.

To further enhance wavefront modulation depth, an effective approach involves the utilization of high-quality hexagonal boron nitride (hBN) encapsulated TMDC monolayers. This offers notable advantages such as enhanced carrier mobility and thus reduced exciton resonance linewidth, consequently leading to a significant enhancement of light-matter interactions [108, 109]. In a related study, Li et al. quantitatively investigated the permittivity and complex refractive index modulation of hBN-encapsulated monolayer MoSe$_2$ across a range of carrier densities, measured when the sample temperature was varied from 4 to 150 K[50]. Three distinct exciton resonances in MoSe$_2$ (A exciton ($X_A$), B exciton ($X_B$), and trion (T) associated with the A exciton) were studied in this work. As shown in Fig. 2c, the refractive index near both the A exciton and B exciton energies shows extremely large gate tunability. At 748 and 754 nm, both the real and imaginary parts can be tuned by over 200% compared to their values at the charge-neutral point. This represents over 5 times larger refractive index modulation than previous studies for both the real and imaginary parts in TMDCs, showing extremely large gate tunability at 4 K. Moreover, by employing an interferometric measurement, applying electrostatic gating leads to 25° phase shift at 754 nm where the change in refractive index is maximal. Notably, such a large voltage-controlled phase shift was achieved in the absence of coupling to antennas or cavities and is limited by material quality (a modest photoluminescence quantum yield, $\eta \sim 25\%$). By integration of gated monolayer $MoSe_2$ with a cavity or metasurface or by improving quantum yield, the phase shift can be increased to >270°. An interesting observation in this study was the appearance of a smooth optically dielectric to metallic phase transition going through an epsilon-near-zero (ENZ) region around the A-exciton (Fig. 2d). This opens opportunities to realize strongly confined and tunable polaritonic modes, which have been explored in other related studies[110, 111].

The large observed index modulation results from the combination of various mechanisms that directly influence the radiative and non-radiative channels, including Coulomb scattering, Pauli-blocking, and screening. First, the screening of the electric field due to increased free carriers weakens the excitonic Coulomb interaction, which reduces the oscillator strength. Second, the Coulomb scattering shortens the coherence lifetime for A exciton and trion which manifests as spectral broadening. Third, Pauli-blocking causes an increase in the optical band gap because of the occupation of lower electronic states. In such a scenario, the A exciton and trion feature blueshift due to a reduction in binding energy as well as an increase in optical gap. By studying the collective effects of these mechanisms in the multi-oscillator system, it is possible to achieve enhanced modulation performance, allowing for better modulation of the exciton response for use in applications such as beam steering[112, 113].

## 3.2. Dynamic beam steering

The remarkable ability to strongly and dynamically manipulate both the amplitude and phase of a light field by harnessing excitonic resonances in TMDCs opens the door to the realization of excitonic phase array metasurfaces for dynamic beam steering. Recently Li et al. demonstrated monolayer molybdenum diselenide $MoSe_2$ excitonic metasurface and one-dimensional reflective-mode dynamic beam steering[107] at a temperature of 6 K. In their device (Fig. 2e), the refractive index of the $MoSe_2$ layer is periodically modulated at the subwavelength scale along $x$-direction by local electrostatic gating of the excitonic resonance. Such a local modulation of the $MoSe_2$'s index with a voltage profile of $V(x)$ in turn, imparts an electrically tunable phase gradient $\phi(x)$ to the reflected wavefront. By varying voltage configurations, the reflected light around the A-exciton resonance (757 nm) can be directed at angles of -30° and 30°, corresponding to the first negative (-1) and first positive (+1) order diffraction patterns, as shown in Fig. 2f. Moreover, the device exhibited a diffraction efficiency of 2.5%. The diffraction efficiency and steering directivity of the device can be further enhanced by using a larger active device area with novel fabrication approaches[114], and by suppression of nonradiative decay mechanisms - using higher quality TMDCs with lower defect density [115] and improved spatial homogeneity[115, 116]. Notably, previous demonstrations of one or two-dimensional spatial light modulators relied on geometric resonances via external cavities [117, 118], this work harnesses the intrinsic cavity-like properties of the exciton itself.

While tunable excitonic resonances in 2-D materials can enable light field modulations at the focusing point and one-dimensional beam steering, programmable two-dimensional beam steering remains challenging since more intricate device geometries are needed. Anderson et al. introduced an elegant method of achieving nanosecond time, 2-D excitonic beam steering using hBN-encapsulated atomically thin $MoSe_2$ reflector[119]. Distinct from 1-D local back gate structures described above, the device by Anderson et al. consists of both top and bottom graphene gates, with the bottom graphene gate covering only a portion of the device (Fig. 3a). Such a split gate configuration enables independent electrostatic doping of the two segments of the $MoSe_2$ channel. As a result, when light is incident on the gate's edge, the two halves of the reflected wavefront gain distinct phases ($\Delta\phi$) and thus constructively interfere in the far-field. Different gate voltage combinations lead to constructive interference at different angles in the far field (Fig. 3b and c).

In the same work, to expand the beam steering into two dimensions, a bilayer $MoSe_2$ is added as a non-resonant reflection source (Fig. 3d). The bottom gate's boundary intersects the dividing line between monolayer and bilayer, enabling control of the relative phase between the three regions. When both the exciton resonances are red-shifted, an upward phase gradient is generated. However, when only one of the monolayer regions experiences this redshift, the phase gradient is directed towards that region (Fig. 3e). With such geometry, the system is compatible with high-frequency beam steering at a switching time down to 1.6 ns. It is also possible to integrate this device into large-scale optical systems. For example, by etching the top and bottom gates into thin perpendicular strips, a 2-D grid of n×n pixels can be realized. To sum up, this fast and highly tunable beam profile modulation platform has shown promising integrability and scalability in even more complex optical systems. The idea of deploying independent channels via gating in two planes can be expanded to encompass intricate beam profile manipulations.

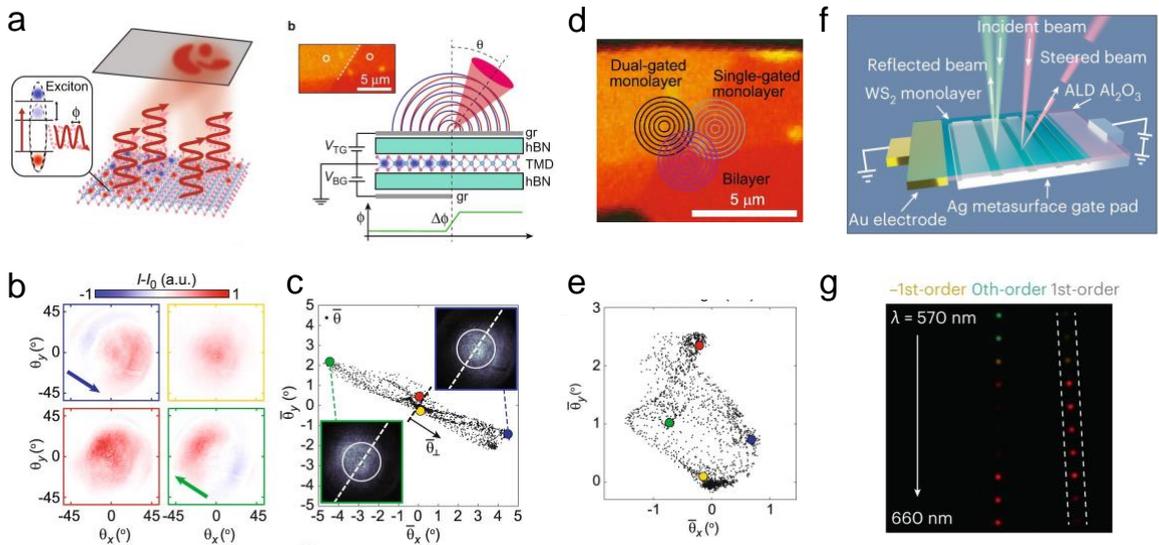

**Figure 3:** (a) Left: schematic of the beam steering device. Continuous tuning of the spatial phase-front is achieved by selective doping of the TMDC using patterned gates, which allows exciton control. Right: schematic of split-gate field effect transistor (SG-FET) structure which allows independent control of optical phase on two sides of the device, resulting in constructive or destructive interference in the far-field. Inset: optical microscope image of such a device, in which the gate edge is indicated by a white dashed line. (b) Fourier-plane imaging of reflected beam ($\lambda_0$ = 755.6 nm) in four different regimes after subtracting the background (which occurs upon high doping ($V_{BG}$=10 V and $V_{TG}$=1.4 V) due to strong quenching of the exciton). A strong beam deflection is observed when the exciton resonance is blue-shifted relative to $\lambda_0$ in only one half of the device (blue, green). However, no deflection is observed if neither half or both halves are blue-shifted (red, yellow). (c) Tracking the center-of-mass beam deflection $(\bar{\theta}_x, \bar{\theta}_y)$ for the full range of gate voltages, showing one-dimensional continuous beam steering. The inset contains Fourier images before any background subtraction. (d) Optical microscope image of the beam-steering device, which shows all three regions - monolayer, bilayer, and gated MoSe$_2$. (e) Tracking the center-of-mass deflection $(\bar{\theta}_x, \bar{\theta}_y)$ demonstrating two-dimensional beam steering. (f) Beam-steering light-field modulator using an active monolayer WS$_2$. A single cell contains a $2\pi$ phase gradient provided by supercells. (g) Back-focal plane imaging of the diffracted light as a function of wavelength from 570 nm to 660 nm in steps of 10 nm. (a–e) are reprinted with permission from Ref.[119], copyright (2022) Springer Nature Limited. (f–g) are reprinted with permission from Ref.[120], copyright (2023) Springer Nature Limited.

Although excitonic resonances in 2-D TMDCs can be harnessed to enhance light-matter interactions and electrical tunability of the optical properties, the strength of exciton resonances is greatly compromised at room temperature due to undesired effects such as dephasing and non-radiative decay processes. The non-radiative mechanisms dominate as both the temperature and carrier density increase. The weak excitonic resonances at room temperature usually lead to a low reflection modulation efficiency of ~0.5%, which is far below the near-unity reflection modulation with TMDC (transition metal dichalcogenides) monolayers at cryogenic temperatures[121]. To overcome this outstanding challenge, one can leverage the Purcell effect to enhance the radiative decay rate of emitters by placing them in a resonant cavity or nanostructures. Employing this theory, Li et al. recently demonstrated a WS$_2$ free-space optical modulator with high reflection modulation efficiency at room temperature[120]. In their device, the WS$_2$ monolayer, which is electrically connected to a gold electrode, is placed on a 15-nm-thick Al$_2$O$_3$ gate oxide with a nanopatterned silver metasurface gate pad underneath it (Fig. 3f). The function of the nanopatterned silver metasurface gate pad is twofold. First, it supports a quasi-guided surface plasmon polariton (SPP) mode around the exciton resonance wavelength (620 nm) of monolayer WS$_2$, which enhances the excitation field and allows the WS$_2$ monolayer absorb and re-emit the resonant light with a higher excitation and radiative decay rate. Second, it serves as a high-conductivity gate pad for charge carrier injection and electrical tuning of the excitonic resonance. Using this device, a 10% reflectance change is observed in the experiment. This corresponds to a 20-fold enhancement compared with modulation using a suspended monolayer in vacuum[106], representing a record for reflection modulation using 2-D excitons in TMDCs at room temperature. By replacing the periodical perturbation with supercells, each providing $2\pi$-phase-gradient, beam steering is also achieved via first-order diffraction, as shown in Fig. 3g.

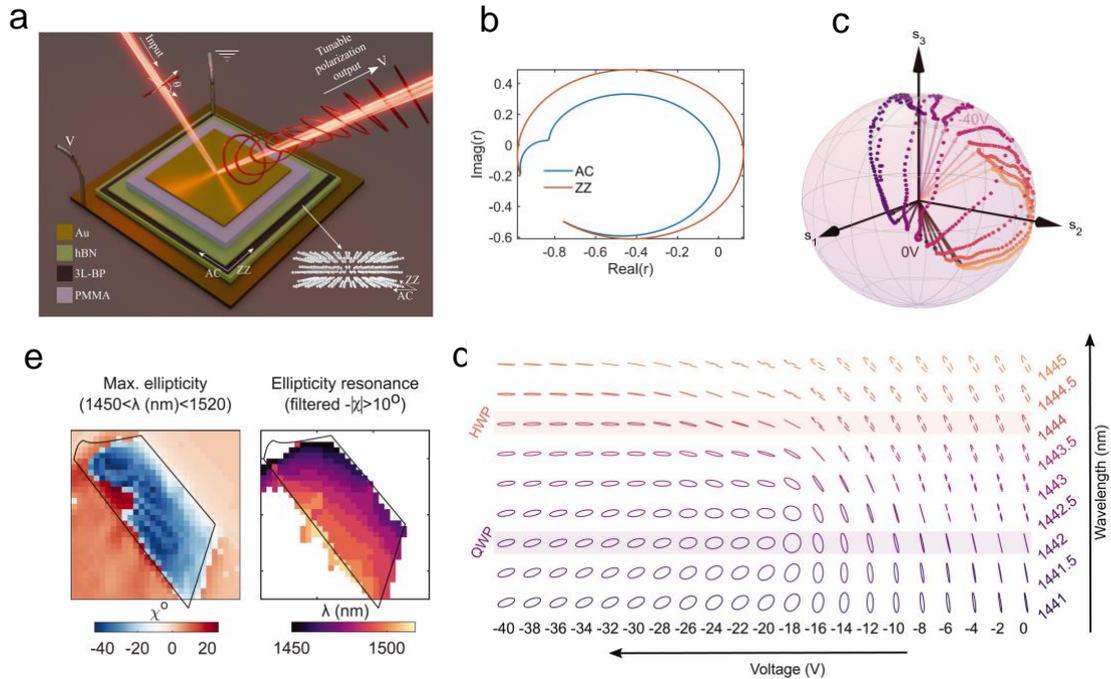

**Figure 4:** (a) Device schematic showing TLBP-cavity enabled polarization conversion. Partially reflective (top) and ideal (bottom) mirrors encapsulate TLBP to convert an incoming linearly polarized beam to different polarization states on the Poincaré sphere for different bias voltages. (b) Broadband anisotropy captured in a theoretically computed complex reflection phasor for a typical cavity structure. A cavity resonance enhanced birefringence is seen ~1480 nm. (c) Selected (different wavelength around cavity resonance) trajectories on the normalized Poincaré sphere for different applied bias voltages showing versatile polarization control. Same color coding as (d). Dark arrows denote zero bias, while light-colored arrows denote the highest bias. (d) Measured reflected polarization ellipse for the same combination of voltages and wavelengths as in (c). Half-wave-plate and quarter-wave-plate-like actions are shown at 1444 and 1442 nm, respectively. Right/left-handedness is shown as solid/dashed lines, respectively. (e) Imaging maximum ellipticity and resonance wavelength in an example cavity-encapsulated TLBP device showing dielectric screening variations. (a–d) are reprinted with permission from Ref.[59], copyright (2021) AAAS.

### 3.3. Active polarization control

While extensive research has been motivated by controlling light's amplitude and phase, active modulation of polarization plays a pivotal role in numerous applications within both classical and quantum information processing realms. One example is the potential enhancement of dichroism sensing and imaging of molecules and biological species[122, 123], as well as performing gate operations on qubits[124], through the use of reconfigurable waveplates and polarizers. Despite the widespread application of polarization modulation, achieving electro-optic polarization control with a high dynamic range of tuning has proven to be challenging, particularly in vdW materials.

Biswas et al. made a significant advancement by showcasing an atomically thin electro-optic polarization modulator operating within the telecom E, S, and C bands (1410-1575 nm wavelength)[59]. Their device comprises a trilayer black phosphorus (TLBP) serving as the tunable birefringent element (Fig. 4a). The selection of the trilayer thickness is guided by the proximity of the highly anisotropic excitonic resonance (~1400 nm) to the telecom band[125]. To expand the dynamic range of birefringence tuning, the TLBP was encapsulated within carefully selected hexagonal boron nitride (hBN) flakes, and gold mirrors were deposited to create a resonant Fabry-Pérot cavity. By adjusting the hBN thickness, the device's working wavelength can be tailored for any point within the aforementioned telecommunication spectral band, as the cavity length determines the optimal operational wavelength. As such, the device exhibits broadband dichroism as seen from the spectral content of the complex reflection phasor plot showing a large difference between the armchair (AC) and zigzag (ZZ) directions (Fig. 4b), with enhancement near the cavity resonance (bump on the AC response) due to stronger light-matter interaction. Electrostatic gates were utilized to control the doping in the TLBP, which effectively quenches excitons by screening, resulting in a reduction of the optical anisotropy between the two principal crystal axes—armchair and zigzag. Here, the cavity is critically coupled via a judicious matching of the cavity and free-space impedance - balancing the different losses in the cavity from the top gold mirror as well as the TLBP rendering extreme sensitivity to small changes in the complex refractive index of the TLBP. Combining spectral and doping-dependent tuning, nearly half the Poincaré sphere is traversed (Fig. 4c). Notably, the device exhibits a tunable quarter/half-wave plate-like behavior at two distinct wavelengths near the cavity resonance, among various other functionalities (Fig. 4d). This result paves the way for further exploration in constructing intricate metasurfaces and stacking multiple BP layers with relative twist angles, enabling complete control of the Poincaré sphere[126]. It also motivates exploration of alternate birefringent systems that hold promise to achieve polarization control in the visible (such as monolayer BP, $ReS_2$)[127, 128] as well as mid- to far-infrared (such as $MoO_3$)[129]. Furthermore, such a device structure combined with Stokes imaging can be used to spatially map birefringence (Fig. 4e) in vdW heterostructures to study Moiré patterns, domains, and grain boundaries as well as dielectric inhomogeneities.

For many sensing applications, it is important to generate and detect very precisely the polarization state of light. It can be expected that the identification of novel birefringent 2-D materials as well as judicious metasurface designs will go hand in hand. A natural follow-up is to explore multi-layer stacking of birefringent materials with different twist angles. Possible applications include the generation and detection of states across the entire Poincare sphere, vortex/vector beam generation, and tuning the topological charge in vortex beams.

## 4. Intelligent sensing
### 4.1. Single-pixel, subwavelength spectroscopy

Spectrometers play a critical role in various scientific research areas and industrial applications. Traditional spectrometers are bulky because they usually consist of movable optical components such as dispersive gratings and interferometers. Miniaturized spectrometers are highly desirable for on-chip applications due to their compact size and their potential in integration with other devices[130]. Most existing on-chip miniaturized spectrometers consist of an array of devices, and each exhibits a distinctive spectral response [82, 131–135]. However, the resolution of these compact spectrometers remains constrained by the number of photodetectors within the array. In addition, their on-chip footprint surpasses the operational wavelength range significantly due to the utilization of numerous photodetectors in this arrangement. Here, we review an emerging spectroscopy scheme based on a single detector by leveraging the strong photo-response reconfigurability of 2-D materials and their heterostructures.

In the mid-infrared wavelength range, Yuan et al. demonstrated an electrically reconfigurable wavelength-scale BP spectrometer, leveraging the bandgap tunability of BP thin film due to QCSE and a spectrum reconstruction algorithm[80]. The device consists of a 13-nm thick BP thin film encapsulated with two hBN thin films capped by monolayer graphene on $SiO_2$/Si substrate, as illustrated in Fig. 5a. The top monolayer graphene and Si substrate function as top and bottom gates, respectively. It has been shown in Ref.[27] that with a displacement field around 1 V $nm^{-1}$, the bandgap of a 10-nm thick BP can be tuned from ~300 to 50 meV (Fig. 5b). Such a bandgap tunability of BP enables the spectroscopy function, as distinctively different spectral responses can be introduced by varying the tuning displacement field. Before characterizing the unknown spectra, it is necessary to determine the spectral responsivity matrix, $R(D, \lambda)$, i.e., the responsivity at known biasing displacement field ($D$) and wavelength ($\lambda$). Such a responsivity matrix can be measured directly, e.g., using a Fourier transform infrared spectrometer (FTIR). In this work, each row of the matrix $R(D, \lambda)$ is inferred by the photo-response vector measured at fixed displacement field $D_i$ under the excitation of multiple light beams with known spectra. Figure 5c plots the measured responsivity matrix $R(D, \lambda)$. The light signal with an unknown spectrum will be measured under a series of displacement fields, generating a photo-response vector, which will be utilized to reconstruct the spectrum together with the known responsivity matrix $R(D, \lambda)$. As shown in Fig. 5d, the reconstructed spectrum derived from a blackbody source aligns closely with the theoretical spectrum dictated by Planck's Law. The active area of the entire device is around 9×16 $\mu m^2$ as illustrated in the inset of Fig. 5d, representing an on-chip spectrometer with a footprint at the scale of the operational wavelength range.

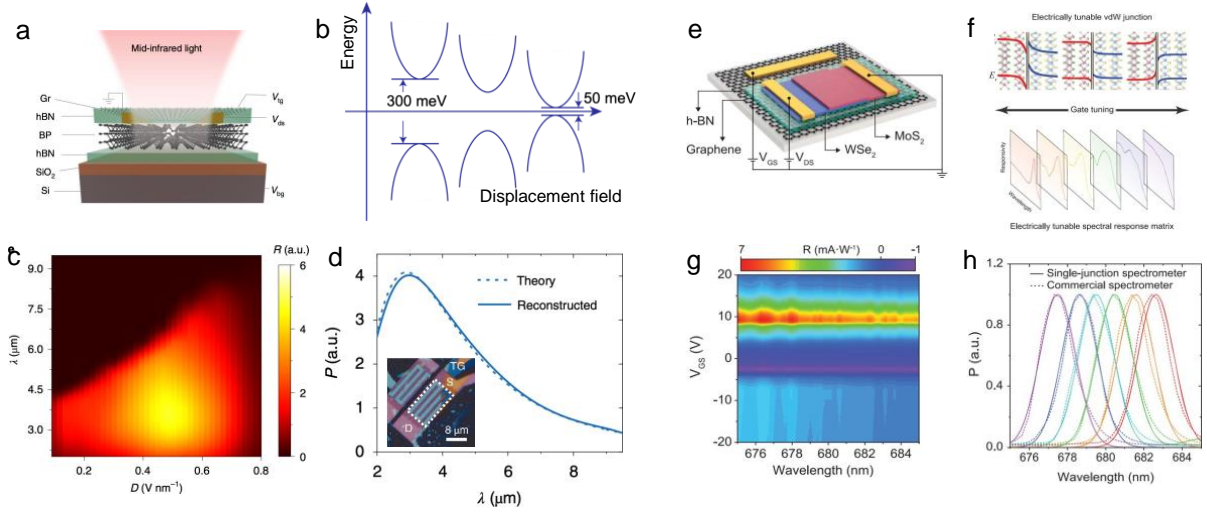

**Figure 5:** (a) Device schematic of the mid-infrared spectrometer based on a single reconfigurable BP device. (b) Illustration of bandgap tuning with applied displacement field. (c) Visualization of responsivity matrix $R(D, \lambda)$. (d) Comparison between reconstructed and theoretical spectra of a black body source. Inset: An optical micrograph of a BP spectrometer. The white dashed rectangular indicates the active area of the spectrometer. (e) Schematic of the $MoS_2$/$WSe_2$ heterojunction spectrometer. (f) Upper panel: illustration of the gate tunable band alignment in the heterojunction. Lower panel: spectral responses under different biasing gate electric fields. (g) Color contour plot of spectral response matrix with 0.1 nm step size. (h) Reconstructed spectra using a heterojunction spectrometer (solid curve) compared to spectra measured by a commercial spectrometer (dashed curve). (a–d) are reprinted with permission from Ref.[80], copyright (2021) Springer Nature Limited. (e–h) are reprinted with permission from Ref.[136], copyright (2022) AAAS.

Beyond bandgap tuning of a single material, the spectral response of the 2-D photodetectors can also be tuned by modulating the band alignment of 2-D heterojunctions. The underlying mechanism behind such a spectral response tuning is complex and multiple factors can contribute to it. To give a few examples, the change of carrier concentration in the junction region will affect the absorption coefficient due to the QCFK effect or the BM shift; the tunable potential barrier for photogenerated carrier transport affects the photocarrier lifetime, thereby influencing the generated photocurrent. Additionally, changes in the depletion region width modify the built-in potential for carrier separation and therefore the generated photocurrent. Importantly, these effects impact photons of different energies differently, leading to spectral response tuning.

In the visible and near-infrared spectral range, Yoon et al. demonstrated a single pixel spectrometer based on a highly tunable $MoS_2$/$WSe_2$ heterojunction photodetector[136] (Fig. 5e). As illustrated in Fig. 5f, in this single-pixel spectrometer, the electro-static gating of the heterojunction can tune the band alignment between $MoS_2$/$WSe_2$, thus affecting interlayer transitions and carrier transport process. Therefore, the absorption spectrum of the device sensitively depends on the gate voltage. In this device, the $MoS_2$/$WSe_2$ heterojunction is encapsulated within two hBN slabs. A monolayer graphene is utilized under the bottom hBN layer as the back gate, and the source and drain electrodes are deposited directly on $MoS_2$ and $WSe_2$, respectively. Photo-response matrix as functions of gate-source bias $V_{GS}$ and wavelength $\lambda$ of this heterostructure spectrometer is measured under the excitation of multiple incident light beams with known spectra. Figure 5g denotes a high-density spectral response matrix ranging from 675 nm to 685 nm with a 0.1 nm learning step. The spectrum of the unknown incident light beam can then be reconstructed by measuring the gate-tunable photo-response vector and the known responsivity matrix. Figure 5h shows that the spectrometer can measure the spectrum of monochromatic light with an accuracy comparable to a commercial spectrometer. Under the 0.1 nm learning step, the average deviation of peak position can be as small as 0.36 nm and two peaks with a 3 nm central wavelength difference can be distinguished. The accuracy and resolution of the system can be further improved

**Table 1**

Comparison of single-pixel spectrometers based on 2-D vdW materials and heterostructures

| Material(s) | Operating Principle | Spectral range($\mu$m) | Spectral resolution(nm) | Footprint($\mu m^2$) | Reference |
|---|---|---|---|---|---|
| BP | QCSE bandgap tuning | 2-8 | ~90 | 144 | [80] |
| BP | QCSE bandgap tuning | 3.07-4.4 | ~25 | ~200 | [81] |
| BP/MoS$_2$ | QCFK and BM | 1.7-3.6 | 43 | 1500 | [67] |
| SnS$_2$/ReSe$_2$ | Band alignment tuning | 0.4-0.8 | 5 | ~300 | [137] |
| BP/MoS$_2$ | Band alignment tuning | 0.5-1.6 | N.A. | 600 | [138] |
| ReS$_2$/Au/WSe$_2$ | Band alignment tuning | 1.15-1.47 | 4.63-4.72 | ~36 | [139] |

with finer learning steps. This single-device spectrometer achieves broadband, high-resolution detection from about 405 nm to 845 nm with a compact footprint of $22 \times 8$ $\mu m^2$.

Besides the two ultra-compact, single-pixel spectrometers mentioned above, several other single-pixel spectroscopy devices have recently been demonstrated based on a similar concept [67, 137–139]. Table 1 summarizes the material choices, operating principles, and device performances of these works. In comparison to spectroscopy approaches relying on a large array of devices with varying spectral responses or external tunable optical components (e.g. tunable filters), the introduction of the material's reconfigurability significantly reduces the number of required devices for spectroscopic tasks, thus minimizing the overall device footprint. Importantly, material tunability also offers flexibility in spectrum measurement processes. For instance, by adjusting operational conditions (e.g., gate biases as seen in the aforementioned devices), the device tuning can be focused on enhancing sensitivity to specific spectral ranges of interest. Such an approach holds the potential to enhance both resolution and measurement speed.

### 4.2. In-sensor neuromorphic optoelectronic sensing

Fueled by advancements in deep learning algorithms which have greatly enhanced image-processing performance, machine vision technology has recently been widely adopted across various sectors including industrial automation, quality control, robotics, autonomous vehicles, medical imaging, and numerous other fields where visual data is crucial for decision-making and operational efficiency. In typical machine vision systems, analog visual data is captured through optoelectronic sensors, converted into digital format, and then analyzed using machine learning algorithms such as artificial neural networks (ANNs). However, the transfer of large volumes of data between optoelectronic sensors and processing units often leads to delays (latency) and high power consumption. With increasing imaging rates and pixel counts, bandwidth constraints pose significant challenges in swiftly transmitting data to centralized or cloud-based computers for real-time processing and decision-making — a critical requirement for time-sensitive applications such as driverless vehicles, robotics, or industrial manufacturing. To tackle this challenge, a promising solution consists of shifting some of the computational tasks, or at least some data pre-processing tasks to the sensory devices at the outer edges of the computer system. This could significantly reduce unnecessary data movement, alleviate the data streaming load to the servers, and improve the bandwidth budget[140, 141]. This architectural paradigm is commonly referred to as in-sensor, near-sensor, or edge computing[142]. In this context, 2-D photodetectors with highly reconfigurable photo-responsivity emerge as promising candidates for building intelligent sensors with in-sensor computing capability – they not only can capture visual information but can also act as artificial neurons to construct artificial neural networks ANN for in-sensor data processing.

Mennel et al. demonstrated a visible image sensor that combines optical sensing with neuromorphic computing[64]. As illustrated in Fig. 6a and b, the image sensor comprises $N$ photoactive pixels organized in a 2-D array, with each pixel subdivided into $M$ subpixels. Each subpixel is a dual-gate WSe$_2$ photodiode (Fig. 6c), whose photo-responsivity can be continuously adjusted via applied dual voltages on the WSe (Fig. 6d). During operation, the optical image projected onto the sensor can be denoted as the input vector $\mathbf{P} = (P_1, P_1, ..., P_N)^T$, where $P_N$ is the optical power incident on each pixel. Sub-pixels of the same color shown in Fig. 6a are interconnected in parallel to generate photocurrent, forming the output vector. As a result, such a device design achieves the matrix-vector multiplication (MVM) operations $\mathbf{I} = \mathbf{RP}$, with $\mathbf{R} = (R_{mn})$ being the photo-responsivity matrix, and $\mathbf{I} = (I_1, I_2, ..., I_M)^T$ being the output vector. The responsivity matrix element $R_{mn}$ can be programmed to be either positive or negative values, depending on the sign of the photocurrent. Thus, the image sensor can simultaneously serve as an ANN capable of both supervised and unsupervised learning tasks, with the synaptic weights encoded within the photo-responsivity matrix through the training process. Using this image sensor, the authors demonstrated two important neuromorphic functions. First, as shown in Fig. 6e left panel, the sensor exhibits classification capabilities, wherein its 3×3 pixel array can categorize an image into one of three classes corresponding to simplified letters, which is a supervised learning task. Second, as shown in Fig. 6e right panel, the sensor can also perform autoencoding. It can generate a simplified representation of a processed image by learning its salient features even in the presence of significant signal noise (Fig. 6f), which is an unsupervised learning task. Remarkably, as shown in Fig. 6g, the sensor achieves an amplifier bandwidth-limited high-speed letter identification within ~50 ns, corresponding to a throughput of 20 million bins per second. Given that the sensor's operation speed is constrained solely by photocurrent generation, this work represents a significant step toward realizing low-latency and ultrafast machine vision.

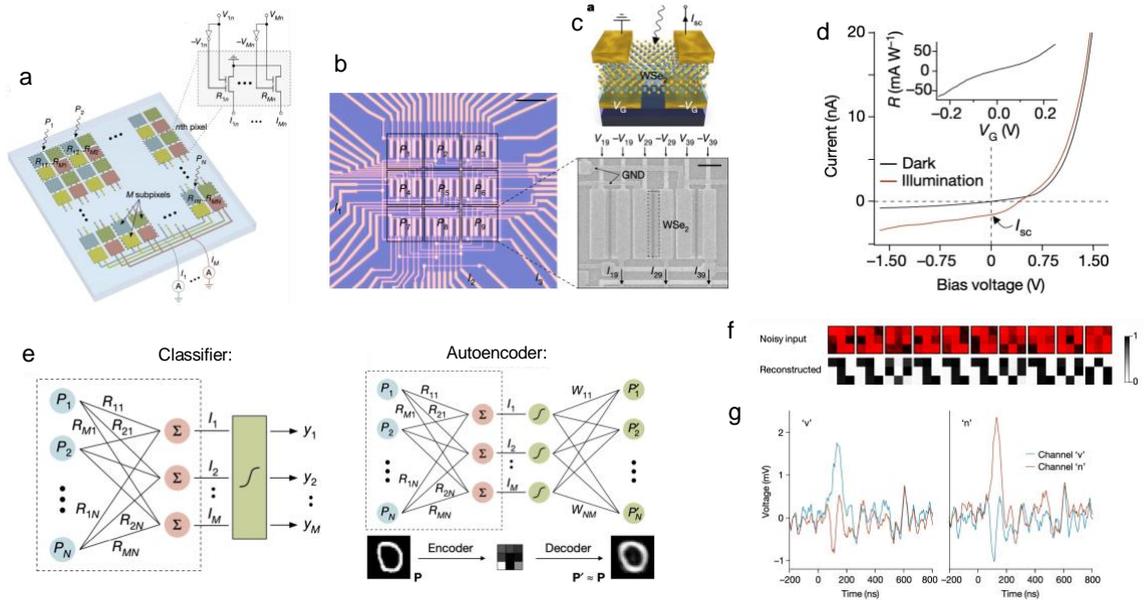

**Figure 6:** (a) The ANN photodiode array. Each pixel is divided into M subpixels. Subpixels of the same color are interconnected in parallel. (b) Microscope image of the photodiode array with 3×3 pixels. (Scale bar, 15 $\mu$m) Inset: Scanning electron microscopy image of one of the pixels. (Scale bar, 3 $\mu$m) (c) Schematic of a single $WSe_2$ photodiode. (d) The current-voltage characteristic curve of one of the photodetectors in the dark (blue line) and under optical illumination (red line). The inset shows the gate-voltage tunability of the photo-responsivity. (e) Schematics of the classifier (left) and the autoencoder (right), below which is an example of encoding/decoding of a 28 × 28-pixel letter from the MNIST handwritten digit database. (f) Randomly chosen noisy inputs and the corresponding reconstructions after autoencoding. (g) Ultrafast image recognition of two different letters 'v' and 'n' with a duration of 40 ns. (a–g) are reprinted with permission from Ref.[64], copyright (2020) Springer Nature Limited.

As the pixel number of machine vision image sensors continues to increase, the need for local non-volatile memory for ANN weight storage becomes imperative. Consequently, an intelligent image sensor with in-memory sensing and computing (IMSC) capabilities is highly desirable[143]. One promising approach to realize this is through the integration of ferroelectric gate dielectrics or floating gate devices into the 2-D photodetectors. These approaches enable the retention of charges in the gate dielectrics, thereby preserving the photodetector's responsivity over extended periods. Along this line, Lee et al. introduced a multi-functional near-infrared image sensor based on a 4 × 3 BP programmable photo-transistors array (bP-PPT array), which has a stack of $Al_2O_3/HfO_2/Al_2O_3$ (AHA) as the gate dielectric and charge storage layer[61], as shown in Fig.7a-c). Notably, such a gate dielectric layer can store high-density charge-trapping sites at ~1.25 eV below the conduction band of $HfO_2$. Consequently, as illustrated in Fig. 1d, by applying electrical voltage pulses of varying amplitudes, the conductance of bP-PPT can be non-volatilely programmed into 5 distinct states (5 digital bits), each corresponding to different photo-responsivity levels. Thus, the bP-PPT array is capable of performing MVM operations, utilizing optical power on each pixel as the input vector, the photocurrent as the output vector, and the responsivity as the weight matrix, akin to the approach demonstrated by Mennel et al[64].

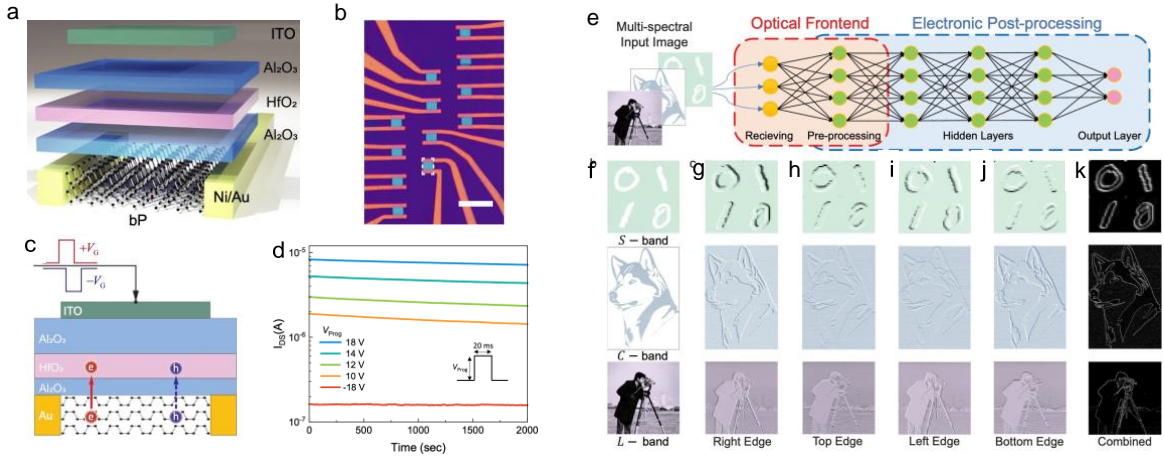

**Figure 7:** (a) The layer structure of a bP-PPT device. (b) Optical microscope image of a 3 ×4 bP-PPT array. (c) Illustration of electrical programming the conductance of a bP-PPT device by gate voltage pulses. (d) The bP-PPT can be electrically programmed to 5 states of conductance levels using pulses of different voltage amplitudes with a fixed pulse duration of 20 ms. A negative repressive pulse (-18 V) can reset the device to the lowest conductance. (e) The architecture of bP-PPT array for in-sensor multi-spectral edge detection. (f) The original input images are encoded in the optical power transmitted in three different telecom bands. (g-j) The resultant images after convolution with the right, top, left, and bottom edge kernels, respectively. (k) The final images combining all the edges. (a–k) are reprinted with permission from Ref.[61], copyright (2022) Springer Nature Limited.

Remarkably, the bP-PPT array showcased by Lee et al.[61] exhibits a programmable photo-response across all telecommunication bands (S, C, and L bands) and the mid-infrared range, rendering it well-suited for multi-spectral in-sensor and in-memory computing at the front-end layer. As illustrated in Fig. 7e, the bP-PPT array receives images encoded in multiple wavelength bands. The array's photo-responsivity matrix is programmed to represent the convolution kernel, enabling direct preprocessing of images in the optoelectronic domain (red dashed line box). Subsequently, the array's conductance matrix is programmed to execute inference computation in the electrical domain (blue dashed line box). Using a 2×2 bP-PPT array, the authors achieved edge detection of 8-bit grayscale images encoded at three telecom bands—1510 nm, 1550 nm, and 1590 nm (Fig. 7f to k). The correlation coefficients between the experimental and simulated results surpass 92% for all three images. The current demonstrated 5-bit programming precision still falls short of that of digital computers, therefore, it prompts exploration into achieving higher programming precision compatible with digital systems and even more intricate deep neural networks. However, the bP-PPT's potential for low-power-consumption and low-latency edge computing has been clearly revealed in this work.

To realize high-performance image perception and recognition using an IMSC architecture based on 2-D photodetectors, photodetectors must enable highly distinguishable, non-volatile weight (photo-responsivity) states with both positive and negative weights for MVM operations. Additionally, it is highly desirable to achieve linear, symmetric, and reversible photo-responsivity tuning characteristics. Wu et al. made important progress toward achieving a near-ideal IMSC device[69]. Their device consists of a 2-D $MoTe_2$ channel with an organic ferroelectric poly(vinylidene fluoride) and trifluoroethylene (P(VDF-TrFE)) as the weight-storage bottom gate dielectric (Fig. 8a). Split local gates (Gate1 and Gate2) beneath the P(VDF-TrFE) dielectric enable independent manipulation of ferroelectric domains on the left and right sides, forming a 2-D $MoTe_2$ homojunction. Remarkably, by applying voltage pulses to the split gates, the ferroelectric domains can be gradually switched. As a result, various junction configurations including p-n, p-p, n-n, and n-p can be formed. This approach allows intermediate photo-responsivity states to be achieved through precise control of the ferroelectric domain configurations and the build-in potentials of the homojunction. Benefiting from such a device physics, 51 (more than five bits) distinguishable photo-responsivity states tunable from negative to positive were obtained (Fig. 8b), which is the most accessible responsivity levels to date. Also, the photo-responsivity changes linearly proportional to the number of voltage pulses from -800 to +800 mA/W (Fig. 8c). Leveraging these impressive capabilities, the authors demonstrated in-situ MVM for simultaneous image sensing and processing. Three distinct image features were experimentally extracted via a one-step convolution using three-in-one kernels. Additionally, the ferroelectric-defined photodiode arrays were trained to perform pattern recognition and classification tasks, enabling a robot dog to be commanded without external memory or computing units, as shown in Fig. 8d.

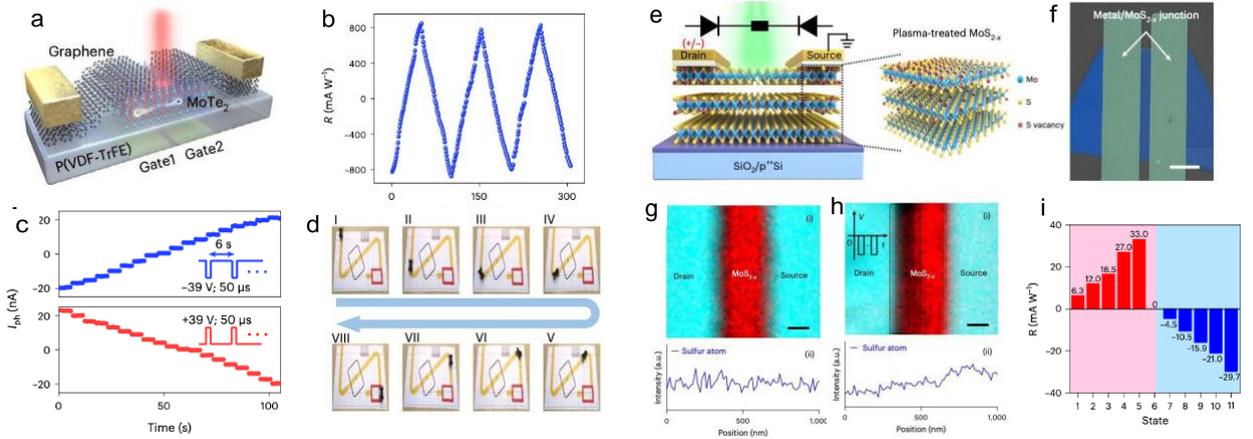

**Figure 8:** (a) Schematic of the reconfigurable 2-D homojunctions with ferroelectric P(VDF-TrFE) as gate dielectric and split local back-gates. (b) Tunable photo-responsivity by ferroelectric domains. (c) Multiple levels of cumulative positive and negative photocurrents with uniform increments and non-overlapping regions. (d) Demonstration of robot dog navigation using the IMSC chip. (e) Schematic of the $MoS_2$ metal/semiconductor/metal (MSM) photovoltaic detector. The $MoS_2$ underneath the electrodes is treated with soft plasma. (f) Scanning electron microscope image of a $MoS_2$ MSM photovoltaic detector. (g) The wavelength-dispersive X-ray spectroscopy (WDS) mapping of sulfur atoms distribution at initial state. (h) The WDS mapping of sulfur atoms distribution after negative pulse programming. (i) Eleven non-volatile responsivity states of plasma-treated $MoS_2$ MSM devices. (a–d) are reprinted with permission from Ref.[69], copyright (2023) Springer Nature Limited. (e–i) are reprinted with permission from Ref.[70], copyright (2023) Springer Nature Limited.

As discussed above, achieving reconfigurable, non-volatile photo-responsivity often requires complex three- or four-terminal device structures, limiting scalability for large pixel arrays. Li et al. proposed an innovative approach to creating simpler two-terminal photodetectors with reconfigurable, non-volatile photo-responsivity[70]. Their device features a simple metal/$MoS_2$/metal architecture, operating as a photovoltaic detector. In the device channel, the $MoS_2$ underneath the electrodes is treated with soft plasma, providing a high density of sulfur vacancies (Fig. 8e and f). By applying voltage pulses to one electrode, they controlled the migration of sulfur vacancies within the $MoS_2$ channel, effectively tuning the local vacancy concentration along the channel, as visualized by the wavelength-dispersive X-ray spectroscopy (WDS) mapping technique (Fig. 8g and h). This process, in turn, modulates the Schottky barrier height at the metal/semiconductor contacts, enabling changes in the polarity and amplitude of short-circuit photocurrents at zero bias voltage. Using this method, the authors achieved 11 distinct, stable positive and negative photo-response states lasting over 1,000 seconds (Fig. 8i). These 11 states, combined with 168 conductance states, were subsequently utilized in a convolutional neural network to demonstrate image processing and object detection.

### 4.3. Bio-inspired optoelectronic sensing

Traditional image sensors excel at capturing static spatial data. However, processing temporal data, such as motion, remains a significant computational challenge, despite its critical applications in autonomous vehicles and surveillance systems. In contrast, biological visual systems—like those in flying insects—can rapidly perceive motion in complex environments, often faster and more efficiently than humans, with much lower energy consumption. This has inspired the development of 2-D optoelectronic volatile devices that can emulate biological visual systems, such as optoelectronic Leaky Integrate-and-Fire (LIF) neurons, graded neurons, optical nociceptors, and multimodal neurons[144]. These devices can exhibit time-dependent conductance changes (volatile states) in response to optoelectronic stimuli. Importantly, these volatile states encode spatio-temporal information, allowing for in-sensor motion perception.

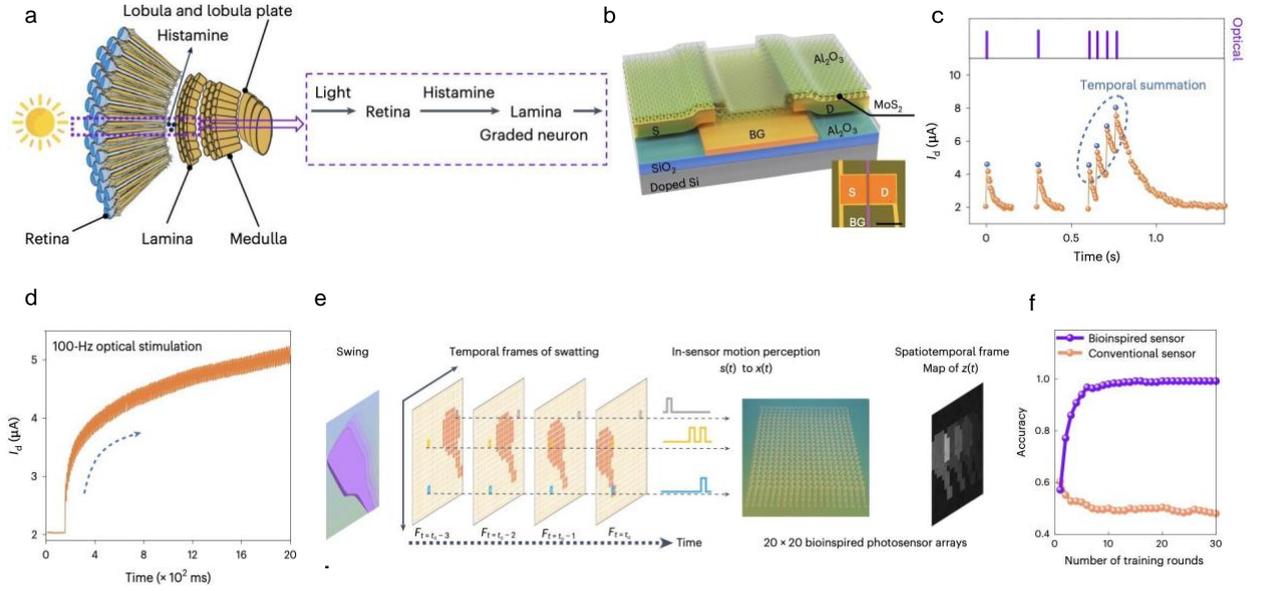

**Figure 9:** (a) The visual system of tiny insects: Lamina neurons are specialized graded neurons responsible for temporal processing. (b) Diagram of the $MoS_2$ phototransistor. Inset: an optical microscopy image showing the phototransistor with labeled source(S), drain (D), and bottom gate (BG) terminals. The scale bar: 200 $\mu$m. (c) Temporal photocurrent response to pulse illumination, showing the occurrence of temporal summation over four consecutive light pulses. (d) Photocurrent response over the illumination time, showing a nonlinear increase followed by saturation. (e) $MoS_2$ phototransistor arrays designed for mapping spatiotemporal visual information. Motion is represented through a sequence of four frames ($F_{t=t_0-3}$, $F_{t=t_0-2}$, $F_{t=t_0-1}$, and $F_{t=t_0}$). Each corresponding column comprises the temporal vision sequence s(t). Bioinspired vision sensors are capable of detecting and converting the temporal sequence s(t) into compressed temporal states, x(t). (f) Accuracy comparison over multiple training rounds for a four-layer neural network utilizing the bioinspired sensor versus a conventional sensor. (a–f) are reprinted with permission from Ref.[145], copyright (2022) Springer Nature Limited.

Inspired by the insect visual system (Fig. 9a), Chen et al. used $MoS_2$ photo-transistors (Fig. 9b) to develop graded neurons for motion perception[145]. In their device, the shallow charge-trapping centers in $MoS_2$ enable graded responses to light stimuli, effectively encoding temporal information directly at sensory terminals, akin to the retina-lamina neural pathway in insects. Under a single 5 ms width light pulse illumination with an intensity of 5 mW/cm$^2$ at 660 nm, the drain current increased dramatically and then decreased to zero due to the release of charges from the shallow trapping centers. A temporal summation appears when the time interval between four sequential pulses is shorter than ∼100 ms due to the incomplete release of charges, as illustrated in Fig. 9c. Furthermore, a nonlinear increase and saturation tendency in the drain current can be observed under pulsed illumination of 100 Hz (∼100 ms time interval between pulses) (Fig. 9 d). Unlike conventional image sensor which only captures spatial information, the 20 × 20 $MoS_2$ phototransistor array (as shown in Fig. 9e) combines the frames to capture spatio-temporal motion information, with distinct photocurrent levels clearly outlining the full trajectory of the motions from left to right, right to left and even approaching or leaving. Combining with a compact four-layer neural network (Fig. 9e), the bioinspired sensor exhibited agile in-sensor motion recognition, achieving over 99% accuracy in tracking object trajectories (Fig. 9f). By modulating the temporal resolution through gate voltage control, the sensor array can perceive motions of varying speeds in nature, from fast-moving objects to slow drifts.

Recent studies also demonstrate that bio-inspired sensors, combining the nonlinear response and decay characteristics (such as short-term and long-term plasticity effects) of 2-D vdW materials, can mimic synaptic behavior in biological systems. These unique characteristics with their photoresponse of 2-D vdW photodetectors can enable more sophisticated information processing at the sensor level. As shown in Fig. 10a, Liu et al.[146] demonstrated an optoelectronic synapse using ferroelectric $\alpha$-$In_2Se_3$, which exhibits controllable relaxation times under electrical and optical stimuli. The device's response to electrical pulses demonstrates short-term potentiation (STP), where the postsynaptic current temporarily increases before decaying, emulating how biological synapses retain short-term memory. This structural configuration also leads to a range of short- and long-term synaptic plasticity behaviors, such as paired-pulse facilitation (PPF) shown in Fig. 10c and paired-pulse depression (PPD) shown in Fig. 10d, which are essential for neuromorphic computation. Comparable dynamics are also observed under optical pulse modulation. Fig. 10e shows that the PPF effect can also be observed by applying a pair of light pulses instead of electrical pulses. Fig. 10f further illustrates two examples on how combining light and electrical stimuli enhances the ability of the synapse to mimic adaptive behaviors of natural neurons, achieving heterosynaptic plasticity. These dynamics enable the $\alpha$-In$_2$Se$_3$ synapse to act as a physical reservoir to process complex multi-modal and multiscale inputs. Fig. 10g demonstrates three different multisensory fusion architectures and their application in a handwritten digit recognition task, in which the left half of the digit is detected through tactile signals due to the coverage of the paint, and the right half through visual signals due to a glass overlay. The highest accuracy of 86.1% is achieved. Moreover, the synapses can also be set in parallel to form a multi-timescale reservoir computer, with different back-gate voltages tuning their relaxation times, as shown in Fig. 10h.

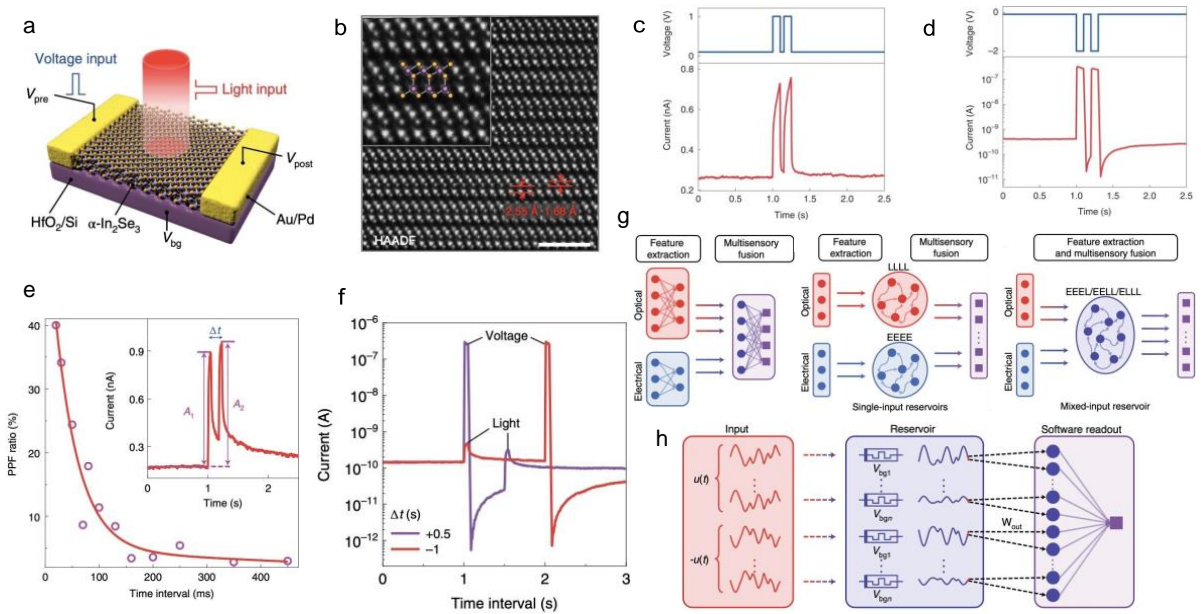

**Figure 10:** (a)The $\alpha$-In$_2$Se$_3$ optoelectronic synapse driven by voltage and light inputs. (b)The crystal structure depicting the dipole locking effect.(c)Paired-pulse facilitation effect, a pair of 1 V, 100 ms electrical pulses are applied with an interval of 50ms. (d)Paired-pulse depression effect, a pair of -2 V, 100 ms electrical pulses are applied with an interval of 100 ms. (e)PPF ratio of different optical pulse intervals $\Delta t$, defined as $(A_2 - A_1) / A_1 \times 100\%$ and fitted by the double exponential decay function in red curve. Inset: The definition of $A_1$, $A_2$ and $\Delta t$, showing the light-induced PPF effect. (f)Two examples of temporal electrical-optical interactions with -0.1 V bias. Electrical pulse: -2 V, 50 ms. Optical pulse: 1.29 mW cm$^{-2}$, 50 ms. (g)Three different multisensory fusion architectures. (h)A multiple-timescale reservoir computing system using optoelectronic synapse as the physical reservoir: inputs are applied to the $\alpha$-In$_2$Se$_3$ devices under different back-gate voltages, resulting in varied relaxation times. (a–h) are reprinted with permission from Ref.[146], copyright (2022) Springer Nature Limited.

These studies highlight the potential for developing intelligent sensing systems that not only detect environmental stimuli but also perform real-time data processing directly at the sensor level using the dynamic optoelectronic responses of 2-D materials. Controlling nonlinear synaptic behaviors—such as paired-pulse facilitation (PPF) and paired-pulse depression (PPD)—opens a promising pathway for creating adaptive optoelectronic neuron arrays that can adjust their responses based on input context. Future advancements in this field could integrate such arrays into autonomous vision systems capable of executing complex tasks like motion prediction, pattern recognition, and environmental adaptation without relying on extensive external computation.

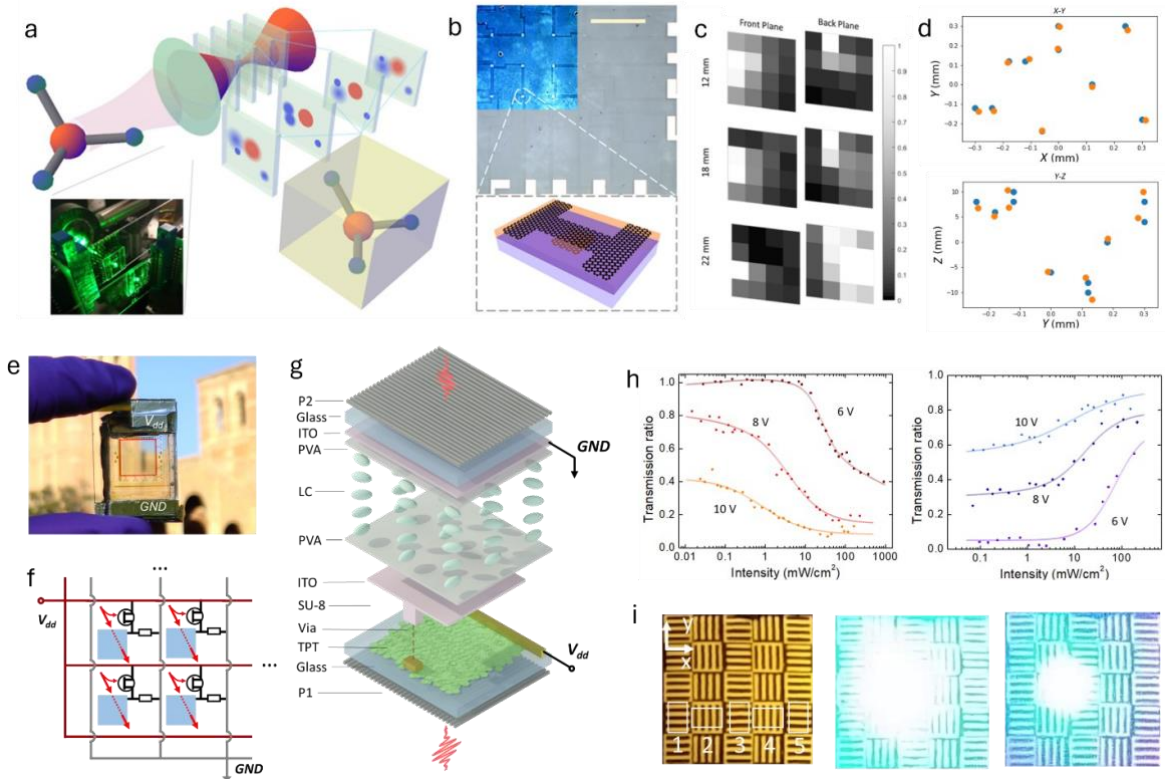

**Figure 11:** (a) The multi-focal plane imaging system enabled by transparent graphene phototransistors. (b) The device structure of graphene floating-gate phototransistor. (c) The raw focal plane images at different point-object depths. (d) Comparison of the predicted (blue) and actual (orange) 3D coordinates of the point object from the focal stack images. (e) Optoelectronic neuron array enabled by $MoS_2$ transparent phototransistors. (f) The electrical layout of the array. (g) The structure of a single optoelectronic neuron pixel. (h) The measured electrically reconfigurable nonlinearity between the input intensity and the transmittance. (i) Implementation of the optoelectronic neuron array for intelligent glare reduction. Left: no glare; middle: with glare but no intelligent glare reduction; right: with intelligent glare reduction, demonstrating selective signal attenuation at glare region while preserving the surrounding object information. (a–d) are reprinted with permission from Ref.[147], copyright (2021) Springer Nature Limited. (e-i) are reprinted with permission from Ref.[148], copyright (2024) Springer Nature Limited.

### 4.4. Transparent intelligent optoelectronic sensing

Another exciting opportunity for intelligent optoelectronic sensing with 2-D vdW materials lies in their high optical transparency due to their atomically thin nature. The high surface-to-volume ratio of vdW materials enables stronger reconfiguration of the material's Fermi level by engineered photo-active heterostructures and charge traps. Hence, it is possible to achieve a large photoconductive gain and responsivity despite the low absorption in a few atom layers[60, 149, 150]. Leveraging this strong photo-gating response and the high optical transparency, one can fabricate sensitive, highly transparent photodetectors based on 2-D vdW materials. Consequently, the mostly transmitted light field can be re-sampled in multiple focal planes for 3-D machine vision. Zhang et al. demonstrated a multi-focal-plane imaging system (Fig. 11a) with highly transparent graphene phototransistor arrays[147]. In the single pixel level, a 6-nm amorphous silicon tunneling barrier is sandwiched by two graphene layers, one as the phototransistor channel and the other as a floating gate. Under illumination, the photocarriers transport across the tunneling barrier and accumulate in the floating gate (Fig. 11b), leading to 3 A/W responsivity with 90% transmittance per 4×4-pixel focal layer. The light field intensities sampled in three dimensions (Fig. 11c) were processed with an external ANN to enable 3D tracking of point sources (Fig. 11d). Further machine learning algorithm studies demonstrated 4-D (positional and angular) light field reconstruction with higher resolution multi-focal-plane transparent photodetector arrays[151, 152].

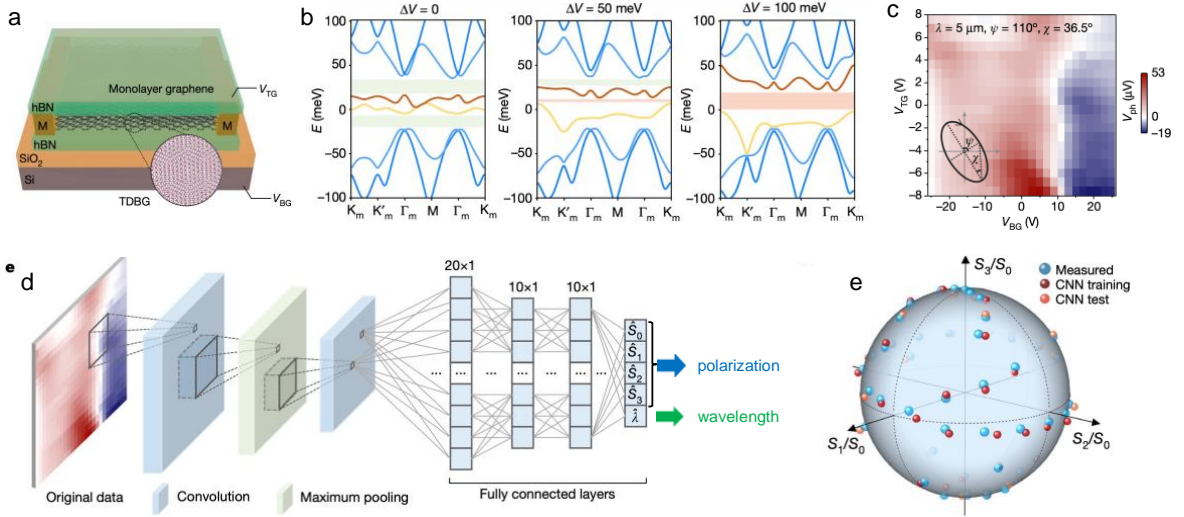

**Figure 12:** (a) Schematic of an intelligent multi-modality sensor made of TDBG with a twisted angle of 1.2°. Inset: schematic of the TDBG moiré superlattice. (b) Calculated band structures of 1.2° TDBG with interlayer potential difference ΔV of 0, 50 meV, and 100 meV. Light green (red) bars denote the moiré superlattice-induced bandgaps. (c) Photovoltage mapping upon the incidence of elliptically polarized light at 5 $\mu$m, with ellipticity and orientation angles $\chi$ = 36.5° and $\psi$ = 110°. The Stokes parameters of the elliptically polarized light are $S_0 = I$, $S_1 = Ip\cos(2\psi)\cos(2\chi)$, $S_2 = Ip\sin(2\psi)\cos(2\chi)$ and $S_3 = Ip\sin(2\chi)$, where $I$ is the incidence light intensity, $p$ is the degree of polarization and $\psi$ and $\chi$ are the orientation and ellipticity angles. (d) Schematic of the convolutional neural networks for Stokes parameters and wavelength inferring. (e) Poincaré sphere showing the polarization states rom the output of the CNN training (red spheres) and test (orange spheres) data sets with their corresponding measured values (blue spheres). (a–e) are reprinted with permission from Ref.[29], copyright (2022) Springer Nature Limited.

The concept of 2-D transparent photodetectors integrated into other functional modules later enabled a detector-modulator stack useful for nonlinear optical computing. Zhang et al. integrated $MoS_2$ phototransistors to liquid crystal modulators (Fig. 11e-g)[148]. The phototransistor's minor (~20%) absorption generates a large photoconductivity change that locally controls the modulator behind. It modulates the transmission of the remaining photons, enabling nonlinear self-modulation of broadband incoherent light with orders of magnitude lower power threshold (56 $\mu$W/cm$^2$) than typical nonlinear optical materials, as shown in Fig. 11h. A 10,000-pixel array was fabricated for intelligent glare reduction that selectively attenuates strong glares while preserving the lower-intensity surrounding object information (Fig. 11i). The optoelectronic neuron array is also desirable for optical computing, such as future optical neural networks directly processing ambient illumination into machine vision results. The exact nonlinearity configurations, such as the threshold intensity and nonlinear contrast, can be reconfigured by changing the voltage applied on the neuron (Fig. 11h), and via electrical doping with an additional gate. These reconfigurable optical nonlinearities could also find opportunities in the optical implementation of novel ANN architectures that depend on trainable nonlinear layers for high performance[153].

### 4.5. Multi-modality optoelectronic sensing

Light is capable of encoding data through its amplitude, phase, wavelength, polarization, and various other properties. Traditional methods of sensing multiple light properties typically require discrete optical components or separate devices. However, in 2-D vdW materials and their heterostructures, certain intrinsic physical properties can simultaneously depend on various properties of light. By fine-tuning these intrinsic physical properties and integrating deep-learning or adaptive algorithms[154], it becomes possible to simultaneously decode multiple physical properties of light using a single device, thereby significantly reducing sensor complexity.

In the work by Ma et al.[29], a tunable graphene moiré superlattice was utilized to capture the rich information of unknown mid-infrared light by measuring the gate-tunable, high-dimensional bulk photovoltaic effect (BPVE). A remarkable characteristic of the BPVE is its pronounced dependence on incident light polarization and wavelength[91]. An excellent material system with a tunable quantum geometry is the twisted double bilayer graphene (TDBG), which exhibits non-zero Berry curvature due to strong symmetry breaking induced by Moiré patterns[155]. Importantly, the berry curvature of TDBG depends not only on the twist angle between individual bilayer graphene, but is highly tunable via the applied out-of-plane displacement field. Building upon this, Ma et al. demonstrated an intelligent sensor comprising a TDBG superlattice at a twisted angle of 1.2°, which is capable of concurrently extracting polarization and wavelength information from light. As depicted in Fig. 12a, in their device, the TDBG superlattice is encapsulated by two hBN slabs, with a monolayer graphene serving as the top-gate electrode and a silicon substrate as the back-gate electrode. As shown in Fig. 12b, the displacement field can alter the complex electronic band structures of TDBG, thus in turn changing the Berry curvature and the BPVE. Upon illumination, the polarization state, power, and wavelength of incident light are encoded into a complex photovoltage map ($V_{ph}(V_{TG}, V_{BG})$) by varying the top-gate ($V_{TG}$) and back-gate ($V_{BG}$) voltages. Figure 12c shows an example of the generated $V_{ph}(V_{TG}, V_{BG})$ map under the incidence of elliptically polarized light at 5 $\mu$m, which can be represented by a polarization ellipse with the ellipticity angle $\chi = 36.5°$ and the orientation angle $\psi = 110°$, as shown in the inset. Subsequently, a properly trained convolutional neural network (CNN) was employed to infer unknown polarization and wavelength information from the map, as illustrated in Fig. 12e. The normalized Stokes parameters ($S_i/S_0$) output by the trained CNN, as depicted in Fig. 12f, show good agreement with measured values at various input polarization states. Moreover, the intelligent sensor can accurately predict incident light wavelengths.

The concept of reconfigurability-enabled multi-modality optoelectronic sensing can also be generalized to traditional material platforms by harnessing different mechanisms. For instance, Tang et al. demonstrated the integration of MEMS with twisted moiré photonic crystals (TMPhC)[154], where moiré physics enables intricate photonic band structures and numerous optical resonances, can be reconfigured via the twist angle and the gap size between the photonic crystal layers. Consequently, employing MEMS-tunable TMPhC alongside an adaptable algorithm allows for simultaneous spectrum and polarization reconstruction within a compact device. Another approach to achieving multi-modality optoelectronic sensing involves optical imaging through spatial and frequency dispersion in thin films[156]. Here, an unknown light beam generates a complex optical image as it traverses through a thin film exhibiting strong spatial and frequency dispersion. This intricate optical information, encompassing spectrum and polarization, is captured by a commercial CMOS image sensor. Subsequently, a deep residue network interprets this image, unlocking its high-dimensional optical content.

## 5. Future perspectives

Looking forward, we envision that reconfigurable 2-D optoelectronics will continue to exert significant influence on optical sensing and information processing. We also expect that 2-D reconfigurable devices will assume new roles in emerging quantum technologies. Here, we outline a few future directions.

**Large-scale reconfigurable 2-D optoelectronics.** The next-generation intelligent optoelectronics will greatly benefit from the recent progress in the large-scale growth of 2-D vdW materials. For example, recently reported centimeter-scale growth of few-layer and thin-film BP[157–159] suggests that it is promising to scale the intelligent sensors and an even larger array of megapixels. The large-scale intelligent sensor array, combined with parallel imaging and programming schemes, such as spatial light modulation and wavelength division multiplexing, can realize more complex deep neural networks for machine vision sensors distributed with edge computing. Integrating large-scale reconfigurable 2-D materials with metasurfaces may enable technologically compelling beam-steering devices with high efficiency as well as large steering angles with higher angular resolution control. It can also enable high numerical aperture lenses for applications in imaging and optical tweezing. In addition, nanophotonic architectures that rely on large unit cells for high quality factors, such as quasi bound states in the continuum (q-BIC)[160] or plasmonic metasurfaces[161]), can be realized in large-scale 2-D materials. Finally, on-chip photonic circuits that often require combining long waveguides, and resonators with large-scale 2-D materials may become more efficient modulators and switches[162, 163].

New tuning knobs for optoelectronic properties. Apart from the electrical tuning methods discussed in this article, we foresee the discovery of tuning mechanisms for the optoelectronic properties of 2-D vdW materials. For example, recent advances have demonstrated that the twist angle between two vdW layers can be continuously adjusted in situ by applying electrical voltages to MEMS structures[154]. This capability enables the continuous tuning of the topological properties of the twisted bilayer, paving the way for the development of novel optical sensors and light sources in both quantum and classical domains.

**Reconfigurable 2-D quantum optoelectronics.** Excitons are expected to continue to play a key role in the 2-D reconfigurable optoelectronics. They can also be harnessed for strong optical nonlinearities when TMDCs are specially/artificially stacked[164] or higher order "Rydberg" excitons are used[112, 165]. An intriguing avenue would be combining 2-D Rydberg excitons with reconfigurability to design transduction elements for converting microwave photons at ~GHz frequencies to optical photons at ~THz frequencies, thereby facilitating the interconnects of quantum hardware. Additionally, excitonic resonances, especially at cryogenic temperatures, can be used as atomically thin optical cavities. For example, two monolayers of $MoSe_2$ can be stacked with a dielectric in between to form resonant Fabry-Perot cavities. Furthermore, thanks to the gate-tunable nature of excitons, the cavity resonance, and losses can be reconfigured on demand, which is crucial for applications such as switchable strong coupling. While most of the 2-D excitons show promise in accessing novel and versatile functionalities, loss remains a challenge. To overcome this, it is important to identify spectral regions where large changes in the real part of the index persist while minimizing the imaginary part.

**Computationally-efficient intelligent sensing.** As deep learning and other intelligent algorithms are expected to play more important roles in advanced optoelectronics, we expect that reconfigurability will become more essential due to the following reasons. First, a reconfigurable device allows for the operation in multiple states, thus significantly reducing the number of devices required by the realization of desirable functions. Second, reconfigurability may improve the efficiency and speed of the measurement processes. For example, in spectroscopy, only a small fraction of the data is useful. Utilization of device reconfigurability may enable a measurement process in which only the most relevant data sets are collected, thus significantly reducing the overhead of the systems. Third, reconfigurability allows for the generation of a large amount of data with a minimal number of devices. These data could be the foundation of the application of different deep learning algorithms.

For in-sensor computing applications, the discussed works achieve fast in-sensor matrix-vector production. However, modern ANNs for computer vision typically exploit a large number of different matrix-vector productions in each layer, and multiple linear/nonlinear layers to achieve good performance. For example, the ResNet typically involves tens to hundreds of convolution layers, each layer with 64 to 512 different reconfigurable kernels[166]. Future improvements demand a higher computation depth in the sensor level without sacrificing the energy and latency advantage. Alternatively, a shallow (one or few linear/nonlinear layers) with a very large number of parallel computation channels may also work as a good encoder for co-designed post-processing algorithms, such as the recent efforts on free-space optical neural networks[167, 168]. Both strategies will demand more functionalities in the reconfigurable sensors as well as efforts on the synergic development of hardware and software architectures.

Furthermore, novel computationally efficient intelligent sensors can be developed to further minimize overhead for sensing systems, such as spectroscopy. In some real-world applications, accurate spectrum reconstruction is unnecessary; what is important is extracting useful information from the spectral data. For example, many remote sensing and hyperspectral imaging applications focus on material identification and classification rather than reconstructing the spectrum. Hence, directly computing the spectral analysis result instead of reconstructing the raw spectra can be a computationally efficient shortcut for such applications.

Another important direction is to further extend the detectable optical parameters for multi-modality sensing. Besides the polarization and wavelength discussed above, detection of optical phase, light field angular distribution, and optical angular momentum with single-pixel or few-pixel detectors were long-sought with various methods. The 2-D vdW material family can provide strongly tunable permittivity and anisotropic photoconductivity, which may enable new devices that can allow optical phase-related parameters to be extracted from reconfigurable intelligent multi-physical sensors.

# CRediT authorship contribution statement

Yu Wang: Writing – review & editing. Dehui Zhang: Writing – review & editing. Yihao Song: Writing – review & editing. Yihao Song: Writing – review & editing. Jea Jung Lee: Writing – review & editing. Souvik Biswas: Writing – figure preparation & review & editing. Fengnian Xia: Writing – review & editing. Qiushi Guo: Writing – figure preparation & review & editing.

## Declaration of Competing Interests

F.X. is a co-inventor of an intelligent sensing patent application in which a trained deep neural network is utilized to interpret high-dimensional photo-response vector/image and another spectroscopy patent application in which tunable materials photo-response and regression algorithm are utilized.

## Acknowledgements

Q.G. and Y. W. acknowledge the support from the Physics Program of the Graduate Center and the Advanced Science Research Center of the City University of New York through the start-up grant. The research activities at Yale University related to this review have received support from the National Science Foundation under grant numbers 1741693 (EFRI NewLAW program) and 2150561, Government of Israel, and Yale Raymond John Wean Foundation.